\documentclass[landscape,12pt,twocolumn]{article}
\usepackage{graphicx}
\usepackage[margin=0.5in]{geometry}
\usepackage{slashed}
\usepackage{dcolumn}
\usepackage{bm}
\usepackage{verbatim}
\usepackage{tabularx}
\usepackage{color}
\usepackage{sectsty}

\sectionfont{\normalsize}

\usepackage{epsfig,amsmath,amssymb,verbatim,mathrsfs,subfigure}

\def\beq{\begin{equation}}
\def\eeq{\end{equation}}
\def\bea{\begin{eqnarray}}
\def\eea{\end{eqnarray}}

\newcommand{\gsim}{ \mathop{}_{\textstyle \sim}^{\textstyle >} }
\newcommand{\lsim}{ \mathop{}_{\textstyle \sim}^{\textstyle <} }

\newcommand{\gev}{\, {\rm GeV}}
\newcommand{\tev}{\, {\rm TeV}}
\newcommand{\mev}{\, {\rm MeV}}

\def\lambdasm{\lambda_{SM}}

\definecolor{red}{rgb}{1,0,0}
\definecolor{lightblue}{rgb}{0.1,0.9,1}
\definecolor{blue}{rgb}{0,0,1}
\definecolor{yellow}{rgb}{1,0.95,0}
\definecolor{pink}{rgb}{0.9,0,1}
\definecolor{green}{rgb}{0.24,0.7,0.28}
\definecolor{darkgreen}{rgb}{0.14,0.6,0.18}
\definecolor{orange}{rgb}{0.9,0.7,0}
\definecolor{grau}{rgb}{0.5,0.5,0.5}
\definecolor{violet}{rgb}{0.8,0.3,0.8}

\begin{document}


\begin{titlepage}
\noindent
\begin{flushright}
{\small CERN-PH-TH-2013-101\\
May 28, 2013}
\end{flushright}

\vspace{0.8cm}

\begin{center}
  \begin{Large}
    \begin{bf}
How well do we need to measure \\ \vspace{0.2cm}the Higgs boson mass and self-coupling?
   \end{bf}
  \end{Large}
\end{center}
\vspace{0.2cm}
\begin{center}
\begin{large}
Rick S. Gupta$^{a}$, Heidi Rzehak$^b$\footnote{On leave from Albert-Ludwigs-Universit\"at Freiburg, 
Physikalisches Institut, Freiburg, Germany.}, James D. Wells$^{b,c}$ \\
  \end{large}
  \vspace{0.6cm}
  \begin{it}

$^{(a)}$IFAE, Universidad Aut\`onoma de Barcelona, 08193 Bellaterra, Spain
\vspace{0.2cm}\\
$^{(b)}$CERN Theoretical Physics, CH-1211 Geneva 23, Switzerland 
\vspace{0.2cm} \\
$^{(c)}$Physics Department, University of Michigan, Ann Arbor, MI  USA 
 \vspace{0.1cm}
\end{it}

\end{center}


\begin{abstract}

Much of the discussion regarding future measurements of the Higgs boson mass and self-coupling is focussed on how well various collider options can do. In this article we ask a physics-based question of how well do we need colliders to measure these quantities to have an impact on discovery of new physics or an impact in how we understand the role of the Higgs boson in nature. We address the question within the framework of the Standard Model and various beyond the Standard Model scenarios, including supersymmetry and theories of composite Higgs bosons. We conclude that the LHC's stated ability to measure the Higgs boson to better than 150 MeV will be as good as we will ever need to know the Higgs boson mass in the foreseeable future. On the other hand, we estimate that the self-coupling will likely need to be measured to better than 20\% to see a deviation from the Standard Model expectation. This is a challenging target for future collider and upgrade scenarios.

\end{abstract}

\vspace{1cm}

\end{titlepage}

\setcounter{page}{2}

\tableofcontents



\section{Introduction\label{sec:intro}}

The ATLAS and CMS experiments have recently announced the discovery of a boson consistent with the Higgs boson having a mass around 126 GeV~\cite{HiggsDiscovery}. The defining property of the Higgs field is that it acquires a vacuum expectation value and  thus leads to  electroweak symmetry breaking (EWSB). In the Standard Model (SM), for instance,  the Higgs potential,
\bea
 V=m^2_{\Phi_{SM}}|\Phi_{SM}|^2 Ê+ \lambda_{SM}|\Phi_{SM}|^4,
 \eea
has a minima at $v \neq 0$. Here, and in the rest of this paper, $\Phi_{SM}$ is the SM  Higgs doublet.  One way to directly check that the discovered boson is indeed a component of the Higgs field responsible for electroweak symmetry is to check the relationship,
\beq
m_h^2 = 2 \lambda_{SM} v^2,
\label{minim}
\eeq
which is obtained by minimizing the above potential and finding the physical Higgs boson ($h$) mass at the minimum. Here $v \simeq 246$ GeV. The above relationship can be experimentally verified by independently measuring the left- and right-hand sides. Whereas the Higgs mass on the left-hand side can be measured by resonant Higgs boson production, the self-coupling on the right-hand side   can be measured by its contribution to the double Higgs production process~\cite{DiHiggs}: $gg \to h^* \to hh$.  

The relationship in eq.~\ref{minim} is of course true only in the SM. In theories beyond the SM (BSM) there can be deviations of the self-coupling from its SM value
$m_h^2/(2 v^2)$. This can happen, for instance, because of the existence of more than one scalar (as in the case of theories with mixed-in singlets and supersymmetric models) or because of the presence of effective higher-dimension operators in the Higgs potential (as in composite Higgs models and  theories with first-order electroweak phase transition). How accurately do we need to measure the left- and right-hand sides of  eq.~\ref{minim}? We use the same criterion that was used in Ref.~\cite{Gupta:2012mi} to answer this question. The self-coupling needs to be measured  at least to an accuracy equal to the maximum allowed deviation in its value (relative to the SM expectation, $\sim m_h^2/(2 v^2)$)   in different BSM theories, assuming no other EWSB states are accessible at the LHC. This is because if other EWSB states are observed we will already know that the Higgs sector is exotic\footnote{In the case of Minimal Supersymmetric Standard Model, we also discuss the deviations depending on the mass of the relevant superpartners to take into account a possible discovery of them.}, whereas if no such states are seen the measurement of the Higgs coupling deviations would be the primary evidence that the Higgs sector is non-standard. We will call this maximum allowed deviation the physics target for the Higgs self-coupling measurement precision. In computing this maximum possible deviation in a given BSM theory, we will also ensure that all existing direct and indirect constraints are  obeyed. 

Regarding the Higgs boson mass, there is a separate reason why we might want to accurately measure it. In some theories the precise value of the Higgs mass  tells us something about UV physics. For instance, it is well known that with a Higgs mass less than $\sim125$~GeV the SM cannot be correct up to arbitrarily high scales because the self-coupling becomes negative below the Planck scale due to renormalization group running.  This  would mean that our universe is in a metastable vacuum. Thus the accuracy of determination of the scale where the self-coupling becomes negative, the so-called triviality  scale, would depend in part on the Higgs boson mass measurement precision.  Another such example is supersymmetry  where the Higgs boson mass is related to the scale of superpartners.  In the first section we will discuss how accurately the Higgs mass needs to be measured to determine the relevant UV physics in these scenarios. In the subsequent sections we compute the Higgs self-coupling measurement precision needed  according to the criterion explained in the previous paragraph in theories with mixed-in singlets,  composite Higgs models, theories with first-order electroweak phase transition, the Minimal Supersymmetric Standard Model (MSSM)  and finally the Next to Minimal Supersymmetric Standard Model (NMSSM). Our conclusion summarizes the results.

\section{How well do we need to measure the Higgs boson mass?}

Measuring the mass of any newly discovered elementary particle is an important scientific endeavor. The ability to test the SM and to test ideas of physics beyond the SM requires us to know them.  The ATLAS and CMS detectors are equipped well to find the precise value of the Higgs boson mass. For a Higgs boson that has  a mass within a few GeV of 126~GeV, one expects through the $H\to \gamma\gamma$ channel to determine its mass to within $0.1\%$\footnote{See for example Fig.\ 10.37 of CMS TDR~\cite{Ball:2007zza} for $30\, {\rm fb}^{-1}$ of integrated luminosity at $\sqrt{s}=14$ TeV. See also Fig.\ 19-45 of the ATLAS TDR~\cite{ATLASTDR}.}, implying $\Delta m_h< 150\mev$.

The conjecture we make here is that there is no conceivable need to measure the Higgs boson mass better than what the LHC can already do.  A proof of that statement is not possible, but we will do the next best thing and show circumstances where it is important to know the Higgs boson mass, and ask if in these circumstances knowing the Higgs boson mass better than 150 MeV would be helpful. In each case the answer is no, but let us follow the details.

In the first illustration, we consider the question of the fate of the universe~\cite{Holthausen:2011aa}. At leading order the Higgs boson self-coupling is related to its mass by $m^2_h=2\lambdasm v^2$. The effective potential that leads to spontaneous symmetry breaking, and therefore our current vacuum state with vacuum expectation value $v=246$ GeV, depends on this coupling as
\beq
V(h)=\frac{\lambdasm(Q)}{4}h^4 +\cdots
\eeq
where the additional terms are small for large field values of $h$. In a renormalization group improved potential the coupling $\lambdasm$ depends on the scale $Q$. If $\lambdasm(Q)<0$ then the Higgs vacuum can be destabilized by the ``turning over" of the potential, and the universe can tunnel to a new catastrophic ground state sometime in the future.  The renormalization group equation for $\lambdasm$ at leading order is~\cite{Wells:2009kq}
\bea
\frac{d\lambdasm}{d\ln Q}& = &\frac{1}{16\pi^2}\left[
\frac{9}{8}\left( \frac{3}{25}g_1^4+\frac{2}{5}g_1^2g_2^2+g_2^4\right) \right. \nonumber \\
& & \left. -\left( \frac{9}{5}g_1^2+9g_2^2+12y_t^2\right)\lambdasm
-6y_t^4+24\lambdasm^2\right],\nonumber
\eea
where $g'$ and $g$ are the gauge couplings  of  $U(1)_Y$ and $SU(2)_L$, and $y_t$ is the top quark Yukawa coupling. The QCD gauge coupling, $g_s$ with $\alpha_s = g_s^2/(4 \pi)$, comes into the renormalization group flow at two loops and also from the one-loop feedback in the top Yukawa coupling flow.

When the Higgs mass is small ($\lambdasm\ll 1$) the $-6y_t^4$ term dominates and tends to push $\lambdasm$ to negative values at higher $Q$, whereas when the Higgs mass is large ($\lambdasm\gsim 1$) the $24\lambdasm^2$ term dominates and tends to increase the value of $\lambdasm$ with scale, keeping the vacuum stable. 
There is an interface between $\lambdasm(Q)$ going negative versus going positive before $Q=M_{Pl}$, and that value determines the critical Higgs mass on this interface. The equation for the critical Higgs mass needed for stability 
is~\cite{Degrassi:2012ry}
\bea
m_h & > &129.4\gev+1.4\gev\left( \frac{m_t-173.2\gev}{0.7\gev}\right) \nonumber \\
& & ~ -0.5\gev\left( \frac{\alpha_s(M_Z)-0.1184}{0.0007}\right)\pm 1\gev
\eea
where $1\gev$ is the theoretical uncertainty, which arises mainly due to the uncertainty in the low-energy matching scale for the Higgs quartic, $\lambdasm$. Given that the Higgs mass is probably between $125$ and $126\gev$, and given the errors expressed in the above equation, it is not totally clear if we are in an unstable or stable vacuum. What is clear is that measuring the Higgs boson mass with accuracy better than the $150\mev$ target that LHC can achieve would be of no help. From direct top quark uncertainties alone,  we would have to know the top quark mass $m_t$
 to much better than $100\mev$, which will not happen given the inherent QCD uncertainties in top quark measurements. At the LHC the best one can hope for is $\delta m_t\sim 1\gev$~\cite{Beneke:2000hk}, and even in the cleaner $e^+e^-$ linear collider environment the best one can hope for is $\delta m_t\sim 100\mev$~\cite{AguilarSaavedra:2001rg} -- a measurement that is a generation away and still not good enough for these purposes\footnote{\label{note1}One could also hope for direct measurement of $y_t$ in $t\bar th$ production at LHC and an $e^+e^-$ linear collider, but this determination accuracy would be well below (see~\cite{Peskin:2012we}) the extractions directly from the top mass measurement assuming the SM.}. Thus, a better measurement of the Higgs mass is of no value for addressing the vacuum stability of the universe question. 

The Higgs boson mass computation in supersymmetry is another example where a precise determination of the Higgs boson mass can be helpful to test and constrain a theory. One can define the Higgs boson mass calculation in the MSSM 
exactly~\cite{Tobe:2002zj} using the definition
\beq
m^2_h=m^2_Z\cos^2 2\beta+\frac{3G_Fm_t^4}{\sqrt{2}\pi^2}\ln\frac{\Delta_S^2}{m^2_t},
\label{susyscale}
\eeq
where $\tan\beta$ is the ratio of vacuum expectation values of the two Higgs doublets of supersymmetry, 
$G_F$ is  the Fermi constant with $v = (2^{1/4} \sqrt{G_F})^{-1} \approx 246$~GeV,
and $\Delta_S^2=m_{\tilde t_1}m_{\tilde t_2}$ is the leading-order radiative correction
 with $m_{\tilde t_1}$ ($m_{\tilde t_2}$) being the lighter (heavier) top squark mass.
 The above expression is subject to threshold corrections, the most important one of these being the correction due to stop mixing (see e.g., \cite{Giudice:2011cg}),
\beq
\delta m_h^2 = \frac{3G_F m_t^4}{2\sqrt{2} \pi^2}\frac{X_t^2}{M_S^2} \left( 1-\frac{X_t^2}{12M_S^2} \right), 
\eeq
where $X_t\equiv A_t-\mu/\tan\beta$ is the stop mixing parameter, and $M^2_S=\tilde m_{t_1}\tilde m_{t_2}$ is the geometric mean of the stop masses. In the language of eq.~\ref{susyscale} this is equivalent to 
\beq
\Delta_S^2=m_{\tilde t_1}m_{\tilde t_2}+m_t^2 \exp \left[\frac{X_t^2}{2 M_S^2}\left(1-\frac{X_t^2}{12 M_S^2}\right)\right].
\label{eq:deltas}
\eeq
There are many unknowns in this equation
at the moment, but if supersymmetry is discovered and the particle content is measured
precisely then a test can be made of the consistency of the prediction for the lightest Higgs
boson mass and the measured value.

To demonstrate how unimportant a more precise measurement of the Higgs mass is to test supersymmetry, let us suppose that $\tan\beta$ is large and known exactly, such that $\cos 2\beta=1$ is a good estimate. Let us then suppose that $m_t$ can be measured to within $100\mev$, which as we have described above will not be possible in the current generation but might be possible in $t\bar t$ threshold scans at a future linear collider. Let us further suppose that $\Delta_S$ can be measured to with $0.5\%$. Even at $1\sigma$ uncertainty this is an extraordinarily optimistic assumption, since $\Delta_S$, taking higher order corrections into account, is a complicated combination of many different superpartner mass and supersymmetry mixing angles. We use $0.5\%$ because that could be the $1\sigma$ statistical (only) error of right-handed squark mass measurements in $e^+e^-\to \tilde q_R\tilde {\bar q}_R\to q\bar q\chi\chi$~\cite{Linssen:2012hp}. The difference between the Higgs mass value when both $m_t$ and $\Delta_S$ are at their central values to both being $1\sigma$ lower than their central values, under these optimistic assumptions, is $\delta m_h=149\mev$, and for $2\sigma$ shifts in $m_t$ and $\Delta_S$ the value is $\delta m_h=297\mev$. This is right near and above the uncertainty in the Higgs mass measurement already anticipated at the LHC, and thus we can conclude that there is no need to measure the Higgs mass to better accuracy for this test within supersymmetry.

Finally, measurement of the Higgs boson self-coupling is a physics measurement goal in next generation experiments. One reason to perform this measurement is to determine if this coupling that governs double Higgs production matches the prediction that one can make for it in the SM which depends directly on the mass of the Higgs boson. Within the SM, the relative uncertainty on the prediction of $\lambdasm$ compared to the uncertainty in the Higgs mass is 
$\delta\lambdasm/\lambdasm \simeq 1.6\%\,(\delta m_h/1\gev)$,
with the coefficient subject to small higher order corrections.  There is no known measurement of the relative Higgs boson self-coupling at the LHC, its upgrades or any $e^+e^-$ collider discussed that could even come close to approaching the sub-percent value needed to test self-consistency with the Higgs boson mass measurement of $150\mev$ anticipated at the LHC.  The best estimates for the LHC upgrades are that $\delta \lambdasm/\lambdasm$ might be measured to within $+30\%$ and $-20\%$ $1\sigma$ accuracy with over $3000\,{\rm fb}^{-1}$ of integrated luminosity at $14\tev$ center of mass energy~\cite{Goertz:2013kp}.

In a subsequent section we will ask a logically separate and more general question of how well do we need to measure $\lambdasm$ for it to be interesting and valuable. The narrow conclusion just above is only that an anticipated measurement of $\lambdasm$ in principle can be compared to the Higgs boson mass measurement as one test of the SM, but given the poor quality of determination of the self-coupling further improvements of the mass measurement would not be helpful. 

The three examples provided here illustrate our general conjecture that the inherent uncertainties in experimental measurements of observables and higher-order computations of theory imply that there is no obvious physics justification for pursuing a Higgs boson mass measurement better than what the LHC can provide.  

\section{How well do we need to measure the Higgs boson self-coupling?}

In this section we will find the maximal self-coupling deviation  from its SM value
 in different BSM theories. Before delving into the BSM scenarios let us first gain our bearings by expanding out the SM Higgs potential to show the relationships between the Higgs boson mass, vacuum expectation value, and self-coupling.  The tree-level BSM scenarios will decouple to the SM results when the BSM decoupling parameters, like supersymmetry breaking mass or compositeness scale, go to infinity.  
 
The SM Higgs potential is
\bea
 V=m^2_{\Phi_{SM}}|\Phi_{SM}|^2 Ê+ \lambda_{SM}|\Phi_{SM}|^4,
 \eea
 which has a minimum at non-zero value of the field when $m^2_{\Phi_{SM}}<0$. In this case one finds
 \beq
 \langle\Phi_{SM}^\dagger\Phi_{SM}\rangle=\frac{v^2}{2},~~~{\rm where}~~v^2=\frac{-m^2_{\Phi_{SM}}}{\lambda_{SM}}.
 \eeq
 At the minimum the Higgs doublet can be expanded about its vacuum expectation value,
 \beq
 \Phi_{SM}=\left( \begin{array}{c} G^\pm \\ \frac{h+v}{\sqrt{2}}+iG^0\end{array}\right),
 \eeq
 where $G^{\pm,0}$ are the Goldstone bosons eaten by the $W^\pm,Z^0$ gauge bosons, and $h$ is the physical propagating Higgs boson.

 The interaction lagrangian of the Higgs boson, when expanded about the minimum is
 \beq
 \Delta {\cal L}=-\frac{m^2_h}{2}h^2-\frac{g_{hhh}}{3!}h^3-\frac{g_{hhhh}}{4!}h^4
 \label{eq:interactionhiggs}
 \eeq
 where 
 \bea
 m_h^2 & \equiv & 2\lambda_{SM}v^2 \\
 g_{hhh} & \equiv & 6\lambda_{SM} v = \frac{3m_h^2}{v} \\
 g_{hhhh} & \equiv & 6\lambda_{SM} = \frac{3m_h^2}{v^2}.
 \eea
In the subsequent sections we will use the normalization convention of eq.~\ref{eq:interactionhiggs} to define what is meant by the three-point ($g_{hhh}$) 
Higgs self-interaction coupling. The Feynman rules for the three point Higgs interaction is
 \bea
hhh:&~~~& -ig_{hhh~} = -3i \frac{m_h^2}{v}.
\eea
Now we discuss the deviations of the $hhh$ couplings from the above value   in three different BSM models.

We should keep in mind that the processes by which the $hhh$ coupling is measured~\cite{Baur:2002qd} at $pp$ and $e^+e^-$ colliders, $gg\to hh$ and $VV\to hh$, can have additional contributions beyond the deviations in the $hhh$ vertex. For example, a heavier Higgs mixing with the lighter Higgs boson can generate an $Hhh$ coupling which can contribute to $gg\to H^*\to hh$. Or, new physics can contribute to the effective gluon-gluon-$h$-$h$ four-point interaction~\cite{Wang:2007zx}. Furthermore, as with the $Hhh$ example, additional dynamics in electroweak symmetry breaking can produce non-SM local vertices in the effective theory $t\bar thh$, $W^+W^-hh$, and $ZZhh$ that contribute to the $hh$ production cross-section. All of these potential extra contributions are not considered, or are taken to be subdominant, in the discussion below. We are interested only in what deviations different theories put on $hhh$ and leave the details of how one measures it, and how it may need to be separated from other dynamics, to another study.

\subsection{Mixed-in singlets}

Let us consider a theory with an extra  singlet where the singlet mixes with the SM Higgs through a renormalizable operator like $|\Phi_{SM}|^2|\Phi_H|^2$~\cite{Wells, Bowen:2007ia}. The lagrangian for our model is,
\bea
 \mathcal{L}_{Higgs} & = &|\mathcal{D}_{\mu}\Phi_{SM}|^2+|\mathcal{D}_{\mu}\Phi_H|^2-
 m^2_{\Phi_{SM}}|\Phi_{SM}|^2 \nonumber\\
 &&Ê-m^2_{\Phi_{H}}|\Phi_{H}|^2-\lambda|\Phi_{SM}|^4
 -\rho|\Phi_H|^4-\eta|\Phi_{SM}|^2|\Phi_H|^2.\nonumber
\eea
The physical component fields are written as
\begin{equation}
\Phi_{SM}=\frac{1}{\sqrt{2}}\left(
\begin{array}{c}
 0\\
  \phi_{SM}+v \\
 \end{array}
\right),~~~
 \Phi_H=\frac{1}{\sqrt{2}}(\phi_H+v')
\end{equation}
where $v$($\simeq 246\gev$) and $v'$ are vevs (vacuum expectation values) around which the $\Phi_{SM}$
and $\Phi_H$ are expanded.  After diagonalizing the mass matrix, we rotate from the
gauge eigenstates ${\phi_{SM},\phi_H}$ to mass eigenstates ${h,H}$.
\begin{eqnarray}
   \phi_{SM}&=&\cos \theta_h h+\sin \theta_h H\\
   \phi_H &=&-\sin \theta_h h+\cos \theta_h H.
   \end{eqnarray}
 The mixing angle $\theta_h$ and the mass eigenvalues are given by
\begin{eqnarray}
\tan \theta_h&=&\frac{\eta vv'}{(-\lambda
v^2+\rho v'^2)+\sqrt{(\lambda v^2-\rho v'^2)^2+\eta^2v^2v'^2}}\nonumber\\
m^{2}_{h,H}&=&(\lambda v^2+\rho v'^2)\pm\sqrt{(\lambda
v^2-\rho v'^2)^2+\eta^2v^2v'^2}.\nonumber
\label{transf1}
\end{eqnarray}
 
Using the above equations we can  express $\lambda$, $\rho$ and $\eta$ in terms of  $m_h^2$, $m_H^2$, $\theta_h$, $v$ and $v'$,
\bea
\lambda &=& \frac{c^2_h m_h^2 + s^2_h m_H^2}{2 v^2} \nonumber\\
\eta&=& c_h s_h \frac{m_H^2-m_h^2}{v v'}\nonumber\\
\rho&=& \frac{c^2_h m_H^2 + s^2_h m_h^2}{2 {v'}^2},
\label{trns}
\eea
where $c_h\equiv \cos\theta_h$ and $s_h\equiv \sin\theta_h$.
The above lagrangian gives the $h^3$ interaction,
\beq
-\left(\lambda c^3_h v -\rho s^3_h v' -\frac{\eta}{2} c^2_h s_h v' +\frac{\eta}{2} c_h s^2_h v\right) h^3.
\eeq
Using eq.~Ê\ref{trns} we can now rewrite the Lagrangian terms for the cubic Higgs interaction as
\beq
g_{hhh}=\frac{3 m_h^2}{ v}\left( c_h^3-s_h^3 \frac{v}{v'}\right).
\label{cub}
\eeq
Recall that the SM value for the Higgs cubic coupling is $3 m_h^2/v$.  

\begin{figure}
\begin{center}
\includegraphics[width=0.8\columnwidth]{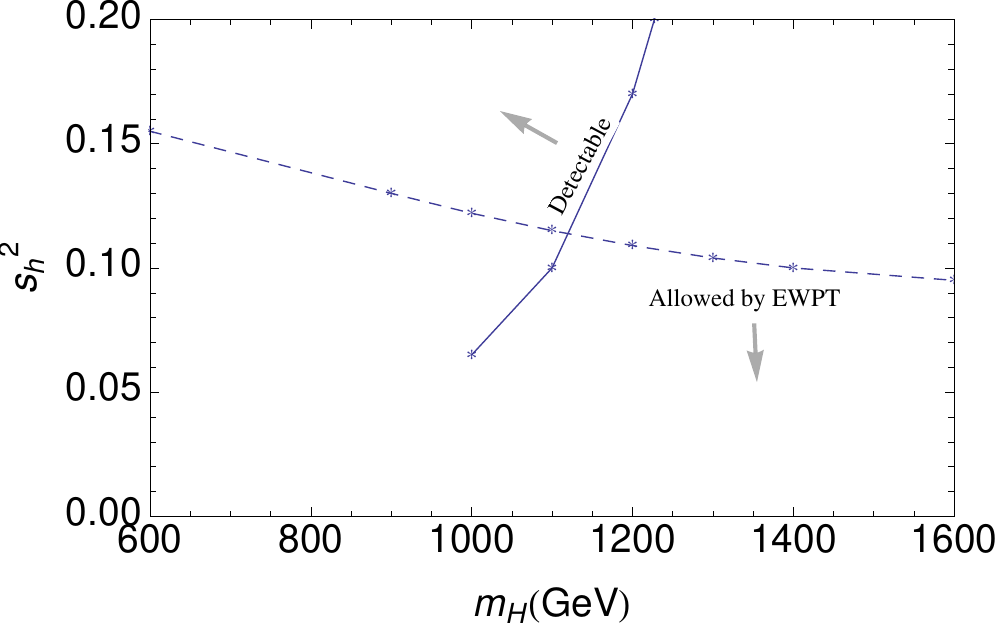}
\end{center}
\caption{The area below the dashed line in the $s_h^2-m_H$ plane is allowed by electroweak precision tests (EWPT) at the 90$\%$ CL in the presence of a mixed-in singlet. The area above, and to the left, of the solid line is the area where the heavy mixed-in singlet Higgs boson is detectable with 100 fb$^{-1}$ data at the 14 TeV LHC.  The maximum allowed $s_h^2$-value that can evade detection is thus given by the intersection of the two lines and is $s_h^2=0.12$. }
\label{fig:singlet}
\end{figure}

The maximum deviation of the tri-Higgs boson coupling from its SM value  occurs for the maximum allowed $s_h^2$ value. This can be found from  Fig.~\ref{fig:singlet}. The region below the dashed curve is the region allowed by electroweak precision tests and the region above the solid curve is the region where $H$ is detectable with 100 fb$^{-1}$ data. For details about how these curves were made see Ref.~\cite{Gupta:2012mi}. Clearly the maximum allowed value of $s_h^2$ given that $H$ is not detectable is
\beq
(s_h^2)_{max}= 12 \%.
\eeq
This tells us that the maximum fractional deviation of the coupling with respect to the SM value eq.~\ref{cub} is,
\beq
 (c_h^3-1)-s_h^3 \frac{v}{v'}~~ \lsim ~ -18\% -4\%\frac{v}{v'} 
\eeq
Although $v'$ is not known, the second term  above is small compared to the first one ($\sim s_h^3$) and can be safely ignored. This gives,
\beq
\frac{\Delta g^{target}_{hhh}}{g^{SM}_{hhh}}= -18 \%.
\eeq
This sets the target for the Higgs boson self-coupling measurement in the context of the mixed-in singlet scenario.

\subsection{Higher-Dimensional operators and composite models}

We want to consider the effect of higher-dimensional operators on Higgs physics. Such theories are low-energy effective theories of a more fundamental theory. The quintessential example of this is the prospect of a composite Higgs boson. That will be our primary focus in this section, with some reference later to another theory of dimension-six operators that gives rise to a possible first-order phase transition for electroweak symmetry breaking. 

Let us consider composite Higgs models where the Higgs is a pseudo-Nambu-Goldstone boson and thus its mass is much lighter than the strong scale. Explicit models realizing this are Little Higgs models~\cite{ArkaniHamed:2002qy} and Holographic composite Higgs models~\cite{Agashe:2004rs}. An effective field theory for such a strongly interacting light Higgs (SILH) has been developed in Ref.~\cite{Giudice:2007fh}. The SILH lagrangian contains higher dimensional operators involving SM fields that supplement the SM lagrangian. It is characterized by two independent parameters: the mass of the new resonances $m_\rho$ and their coupling $g_\rho$. The ``pion'' decay constant $f$ is given by,
\beq
m_\rho=g_\rho f
\eeq
where $g_\rho \leq4 \pi$.

Here we do not list all the operators in the SILH lagrangian but only those relevant to us, i.e.\ those that affect the triple Higgs coupling:
\begin{equation*}
{\cal L}_{SILH}  =  \frac{c_H}{2f^2} \partial^\mu (\Phi_{SM}^\dagger \Phi_{SM}) \partial_\mu  (\Phi_{SM}^\dagger \Phi_{SM}) +\frac{c_6 \lambda}{f^2} (\Phi_{SM}^\dagger \Phi_{SM})^3.
\label{silh}
\end{equation*}
The coefficients of the above operators have been estimated using Naive Dimensional Analysis~\cite{Manohar:1983md, Giudice:2007fh} such that the couplings $c_H$ and $c_6$ are expected to be ${\cal O}(1)$ numbers. Electroweak precision constraint from the  measurement of the $S$-parameter\footnote{Note that the constraint from the $T$-parameter is more severe but this is avoided by imposing custodial symmetry in specific  composite Higgs models.}  requires at 90\%~CL~(see \cite{Giudice:2007fh} and~\cite{Pappadopulo:2013vca}),
\beq
m_\rho \gtrsim 2.6 {~Ê\rm TeV.}
\label{es1}
\eeq
Precision constraints also require at 90\%~CL~\cite{Gupta:2012mi},
\beq
c_H \xi \lesssim 0.15,
\label{es2}
\eeq
where $\xi = v^2/f^2$. 

Let us now review how to obtain the triple Higgs coupling expression in the presence of these additional operators. The Higgs potential is modified by the operator with coupling $c_6$ as follows,
\begin{equation*}
V(H)= \mu^2 (\Phi_{SM}^\dagger \Phi_{SM}) + \lambda (\Phi_{SM}^\dagger \Phi_{SM})^2+\frac{c_6 \lambda}{f^2} (\Phi_{SM}^\dagger \Phi_{SM})^3.
\end{equation*}
We can eliminate $\mu^2$ using the minimization condition,
\beq
\mu^2=- \lambda v^2- \frac{3}{4} c_6 \lambda \xi v^2 
\label{muu}
\eeq
and then find the mass of the Higgs boson at the minima,
\beq
m_h^2=2 \lambda v^2 + 3 c_6 \lambda \xi v^2 
\label{lamba}
\eeq
where we have used eq.~\ref{muu}. 

Now let us write down the cubic Higgs terms from the SM and the operators above,
\beq
\left(-\lambda -\frac{5}{2} c_6 \lambda \xi\right) h^3v. 
\eeq
Substituting $\lambda$ from eq.~\ref{lamba} results in,
\beq
-\frac{m_h^2}{2 v^2}(1+c_6 \xi) v h^3.
\eeq
Finally we must consider the effect of the operator proportional to $c_H$ in eq.~\ref{silh} which leads to the following derivative couplings,
\beq
\frac{c_H \xi}{2}\left(1+\frac{h}{v}\right)^2 \partial_\mu h \partial^\mu h
\eeq
These terms can be eliminated by the following redefinition of the Higgs field~\cite{Giudice:2007fh},
\beq
h \to h- \frac{c_H \xi}{2} h \left( 1 + \frac{h}{v} + \frac{h^2}{3 v^2}  \right ).
\eeq
Note that the above redefinition shifts $m_h$ but not the Higgs vev. Thus we recover the expression for the triple Higgs coupling, $g_{hhh}/3!$,  correct up to order $\xi$, already presented in Ref.~\cite{Giudice:2007fh},
\beq
-\frac{m_h^2}{2 v}\left(1+c_6 \xi-\frac{3}{2} c_H \xi\right)  h^3. 
\eeq
In the above equation $\xi \to 0$ is the SM limit. Using the constraint in eq.~\ref{es2} we can thus derive the self-coupling target,
\begin{equation*}
\frac{\Delta g_{hhh}}{g_{hhh}}=\frac{c_6}{c_H}  (c_H \xi)_{max}-\frac{3}{2} (c_H \xi)_{max}\approx 0.15\frac{c_6}{c_H}  -0.23.
\label{eq:ghhh1}
\end{equation*}

In the past in a different context  with a different  theoretical motivation,
the operator $(\Phi_{SM}^\dagger \Phi_{SM})^3$ has been studied.
With such an operator one can achieve a first-order electroweak phase transition~Ê\cite{Delaunay:2007wb} that may be relevant for the problem of baryogenesis. It was shown that one needs
\beq
\frac{f}{\sqrt{c_6}} >550 {\rm~GeV}
\eeq
to ensure that the expansion of the universe does not disallow percolation of bubbles of metastable vacuum. This corresponds to $c_6 \xi<0.2$. This would mean that the maximum coupling deviation allowed in such a scenario would be,
\bea
\frac{\Delta g_{hhh}}{g_{hhh}}&=&c_6 \xi = 0.2.
\label{eq:ghhh2}
\eea

Equations~\ref{eq:ghhh1} and~\ref{eq:ghhh2} set a target of $\sim 20\%$ needed for measuring the Higgs boson self-coupling interaction in these higher dimensional operator theories, including composite Higgs models and models with Higgs potentials that allow first-order electroweak phase transitions.

\subsection{Minimal Supersymmetric Standard Model}
\label{sec:mssm}

The Minimal Supersymmetric Standard Model (MSSM) exhibits an extended Higgs sector with two Higgs boson doublets, $H_d$ and $H_u$, which couple to down- and up-type quarks, respectively. The two neutral, CP-even components of the Higgs boson doublets $H_d^0$, $H_u^0$ mix and form the mass eigenstates $H$ and $h$. For our discussion, we will assume that the lighter CP-even Higgs boson $h$ is the SM-like one. 
In addition to the Higgs sector extension with respect to the SM, the MSSM is comprised of superpartners to the particles of the two Higgs doublet model which might be discovered at the LHC depending on the scenario realized in nature. For the investigation of the Higgs sector, the top squarks, superpartners of the top quarks, are most relevant. 

In the following, we will discuss the maximal possible deviation of the MSSM triple Higgs coupling with respect to the SM one if no other Higgs boson will be discovered at the LHC. We model the LHC discovery potential according to Fig.~1.21 of~\cite{Linssen:2012hp} inspired from chapter~19 in~\cite{ATLASTDR}. To take into account the possible discovery of superpartners at the LHC we will discuss the maximal deviations, noting its dependence of the mass of the most relevant superpartners, the top squarks.

At tree-level the tri-Higgs boson coupling is
\beq
g^{tree}_{hhh}=  \frac{3 m_Z^2}{v} \cos (2\alpha) \sin (\alpha+\beta).
\label{eq:ghhhtree}
\eeq
with $v = (2^{1/4} \sqrt{G_F})^{-1} \approx 246$~GeV and $G_F$ being the Fermi constant. The angle $\alpha$ describes the transformation from the Higgs boson interaction eigenstates to the mass eigenstates. The $Z$~boson mass is denoted by $m_Z$ and the angle $\beta$ is defined as the ratio of the Higgs vacuum expectation values, $\tan \beta = v_u/v_d$. In the decoupling limit of $m_A\to \infty$ one finds that  $\alpha \to \beta - \pi/2$ and $m_h^2=m_Z^2\cos^2 2\beta$. Substituting this into eq.~\ref{eq:ghhhtree} gives $g^{tree}_{hhh}\to 3m_h^2/v$, matching the SM result as expected.

At tree-level, the mass of the lighter CP-even Higgs boson has an upper theoretical bound at $m_Z$; however, large radiative corrections shift the mass to higher values. In order to compare the triple Higgs coupling in the MSSM and the SM,  it is important to apply the same approximations in both cases -- which includes calculating the Higgs mass and the triple Higgs coupling to the same order. It is clear that we have to take into account higher order effects. However, we cannot make use of all the higher-order corrections known for the computation of the mass of the lightest MSSM Higgs boson when the analogous corrections for the triple Higgs coupling are not known.

In Ref.~\cite{llog} renormalization-group improved corrections to the effective potential are presented, taking into account dominant one- and two-loop contributions. For our investigation, using this effective potential and deriving the Higgs mass and the triple coupling from there has the advantage of automatically applying the same approximation for the Higgs mass and the triple Higgs coupling. Also, the two-loop contributions have a sizeable effect on the coupling deviations. See also the discussion in the appendix with results of various other more limiting approximations.

The effective potential, used in Ref.~\cite{llog}, is defined according to 
the conventions of Ref.~\cite{HaHe} as
\bea
V^{\text{eff.}} &= & \frac{\lambda_1}{2} |H_d|^4 + \frac{\lambda_2}{2} |H_u|^4 
+ \lambda_3 |H_d|^2 |H_u|^2+ \lambda_4 |H_d^\dagger H_u|^2  \nonumber \\
& &  + \left[ \frac{\lambda_5} {2}|H_d^\dagger H_u|^2 +\left(\lambda_6 |H_d|^2 + \lambda_7  |H_u|^2\right) H_d^\dagger H_u + c.c\right], \nonumber
\label{VeffMSSM}
\eea
where the coefficients $\lambda_i$, $i = 1,\dots,7$ contain also the loop corrections. 
Explicit expressions for the coefficients  $\lambda_i$ can be found in the appendix.
With $H_d$ and $H_u$ developing a vacuum expectation value, 
the effective potential can be expanded about these vacuum expectation values, and the triple Higgs vertices in terms of interaction eigenstates 
can be derived:
\begin{align}
g_{H_d^0H_d^0H_d^0} &= 
3 v[\lambda_1 \cos \beta + \lambda_6 \sin \beta]~,\\
g_{H_d^0H_d^0H_u^0} &= 
 v [3 \lambda_6 \cos \beta + (\lambda_3+\lambda_4+\lambda_5) \sin \beta]~,\\
g_{H_d^0H_u^0H_u^0} &= 
 v [3 \lambda_7 \sin \beta + (\lambda_3+\lambda_4+\lambda_5) \cos \beta]~,\\
g_{H_u^0H_u^0H_u^0} &= 
 3  v [\lambda_2 \sin \beta + \lambda_7 \cos \beta]~.
\end{align} 
The transformation into mass eigenstates can be done with the help of the effective mixing angle $\alpha$, obtained by diagonalizing the Higgs mass matrix while including the corresponding higher-order corrections. This leads to
\begin{align} \nonumber
g_{hhh} &= - \sin^3 \alpha g_{H_d^0H_d^0H_d^0} + 3 \sin^2 \alpha \cos \alpha g_{H_d^0H_d^0H_u^0} \\& \quad 
 - 3 \cos^2 \alpha \sin \alpha g_{H_d^0H_u^0H_u^0} 
+ \cos^3 \alpha g_{H_u^0H_u^0H_u^0}~. \label{ghhh}
\end{align}

In the decoupling limit,
 the above potential gives a Higgs mass, 
 \begin{align}\nonumber
m_h^2 &= v^2\bigl[\lambda_1  \cos^4\beta + \lambda_2  \sin^4 \beta 
+ 2 \bigl(\lambda_3 + \lambda_4 + \lambda_5\bigr) \sin^2 \beta \cos^2 \beta  \\&\qquad 
+ 4 \sin \beta \cos \beta\bigl(\lambda_6 \cos^2 \beta +  \lambda_7 \sin^2 \beta\bigr)\bigr]~, 
 \end{align}
and the triple Higgs coupling reduces to the correct SM value.

The coefficients $\lambda_i$ given in Ref.~\cite{llog} comprise the dominant one-loop $\mathcal O(y_t^4)$ contributions, stemming from the large 
top Yukawa coupling $y_t$, as well as one-loop corrections of order $\mathcal O(m_Z^2/v^2 y_t^2)$, $\mathcal O(y_t^2y_b^2)$, $\mathcal O(y_b^4)$ and
 $\mathcal O(m_Z^2/v^2 y_b^2)$ where the latter ones are proportional to the bottom Yukawa coupling $y_b$. The dominant two-loop contributions are 
of the order $\mathcal O(y_t^4 \alpha_s)$ 
and $\mathcal O(y_t^6)$. They are included into the $\lambda_i$ as well as two-loop terms proportional to $y_b$ of order
 $\mathcal O(y_b^4 \alpha_s)$, 
$\mathcal O(y_b^6)$, $\mathcal O(y_t^4y_b^2)$ and $\mathcal O(y_b^4y_t^2)$.

For the investigation of the deviation of the triple Higgs coupling from the SM one we performed a scan over several parameters: 
$\tan \beta$ from 2 to 45 and
 the CP-odd Higgs boson mass $m_A$ from 200~GeV to 800~GeV, which are the free parameters in the Higgs sector at tree-level, 
 the diagonal soft breaking parameters of the stop mass matrices $M_{L_{\tilde{Q}_3}} = 
M_{R_{\tilde{t}}}$ from 100 GeV to 3 TeV, the Higgs superfield mixing parameter $\mu$ between $\pm 1$~TeV and the stop mixing parameter 
$X_t = A_t - \mu/\tan \beta$ between $\pm 150~\text{GeV}\cdot n_{\text{max}}$, where $n_{\text{max}}$ is the nearest bigger integer to 
$\sqrt{6} M_{L_{\tilde{Q}_3}}/(150~\text{GeV})$ ($A_t$ being the top soft breaking trilinear coupling).

The top Yukawa coupling is determined as $y_t = \sqrt{2} \overline{m}_t/(v \sin \beta)$ with the running top quark mass 
$\overline{m}_t =  m_t/[1+4 \alpha_s/(3 \pi)]$, $m_t  = 173.2$~GeV.  The bottom Yukawa coupling is 
\begin{align}
y_b = \frac{\sqrt{2}}{v \cos \beta}\frac{\overline{m}_b}{1+\Delta_b}
\end{align}
where~\cite{Carena:2000uj}
\begin{equation*}
\Delta_b = \frac{2 \alpha_s}{3 \pi} m_{\tilde{g}} \mu \tan \beta
                                    I(m_{\tilde{b}_1},m_{\tilde{b}_2},m_{\tilde{g}}) 
                                      +\frac{y_t^2}{16 \pi^2} A_t \mu \tan \beta
                                 I(m_{\tilde{t}_1},m_{\tilde{t}_2},\mu).
\end{equation*}
The function $I$ is defined as
\begin{equation*}
I(m_1,m_2,m_3) = -\frac{m_1^2 m_2^2 \log \frac{m_1^2}{m_2^2}
              +m_2^2 m_3^2 \log\frac{m_2^2}{m_3^2}
               +m_3^2 m_1^2 \log\frac{m_3^2}{m_1^2}}
               {(m_1^2-m_2^2) (m_2^2-m_3^2)(m_3^2-m_1^2)}.
\end{equation*}
The running bottom quark mass is chosen as $\overline{m}_b = 2.9$~GeV, the gluino mass as $m_{\tilde{g}} = 1$~TeV and the diagonal right-handed 
soft-SUSY-breaking parameter in the sbottom-mass matrix as $M_{\tilde{b}} = 1.2$~TeV. The resulting sbottom masses are denoted as $m_{\tilde{b}_1}$,
  $m_{\tilde{b}_2}$.

\begin{figure*}
\begin{center}                          
\includegraphics[width=0.9\columnwidth]{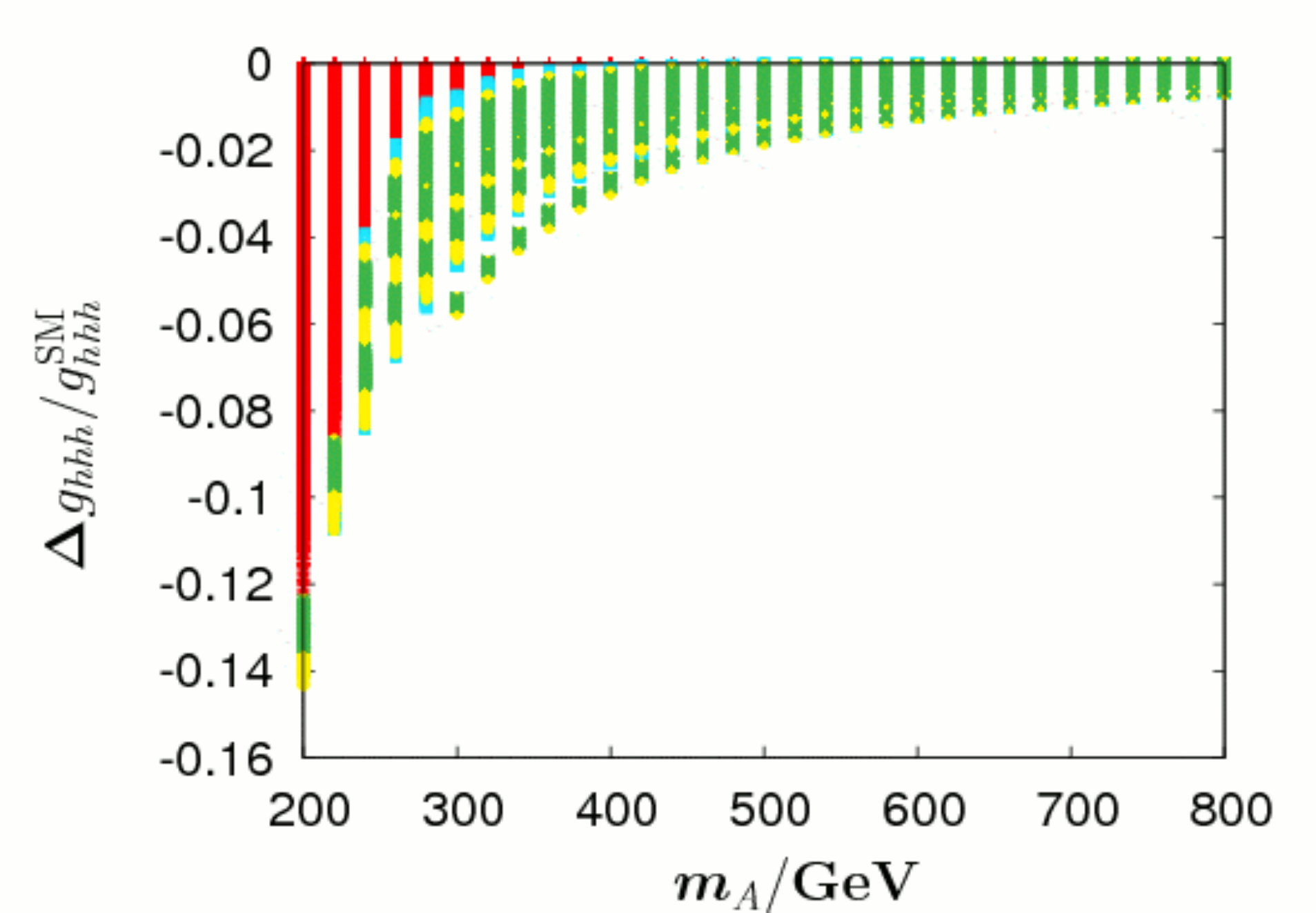}
\includegraphics[width=0.9\columnwidth]{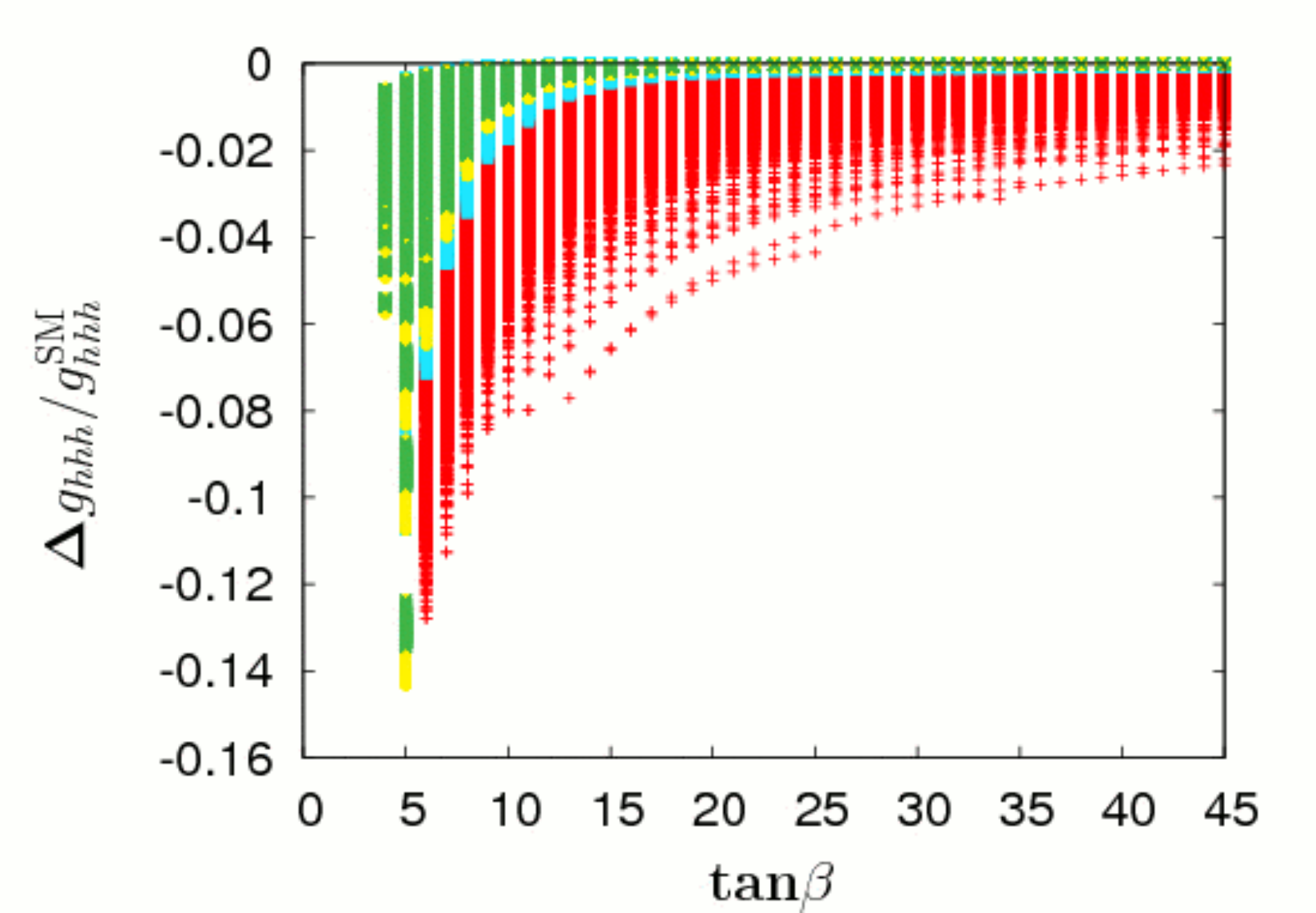}
\end{center}
\caption{The triple Higgs coupling deviations in the MSSM as a function of $m_A$ in the renormalization-group 
improved leading-log approximation, with $y_t$ defined via the running top quark mass.
The red ({\color{red}  $\boldsymbol +$}) region 
corresponds to points in the $\tan \beta - m_A$ plane lying in the several Higgs boson discovery region, while all the 
other points lie in the single Higgs boson discovery region. 
The lightblue ({\color{lightblue}$\blacksquare$}), yellow ({\color{yellow}$\bullet$}) and green 
({\color{green}$\boldsymbol \times$})  points correspond to mass values of the lighter stop of $m_{\tilde{t}_1}< 1.0$~TeV, the  
 $1.0~\text{TeV} \le m_{\tilde{t}_1} < 2.5$~TeV and  $m_{\tilde{t}_1} \ge 2.5$~TeV, respectively. Plot on right is same but versus 
$\tan\beta$.} 
\label{fig:hhhMSSM2lfull}
\end{figure*}

The deviations of the triple Higgs coupling,  $\Delta g_{hhh}/g^{SM}_{hhh}$ with $\Delta g_{hhh} = g_{hhh} - g^{SM}_{hhh}$, are shown in Fig.~\ref{fig:hhhMSSM2lfull} versus $m_A$ and $\tan \beta$ in the left and the right plot, respectively. The deviations are calculated taken into account all the corrections given in Ref.~\cite{llog}. 
The parameter points either fulfill the mass constraint $122~\text{GeV}\le m_h\le 129~\text{GeV}$ or a relaxed mass constraint. In the case of the relaxed mass constraint the Higgs boson mass is calculated neglecting the terms $\mathcal O(m_Z^2/v^2 y_t^2)$. If the resulting Higgs mass fulfills the constraint $122~\text{GeV}\le m_h\le 129~\text{GeV}$ the parameter point is kept (the deviations however are calculated with the full expressions).  The terms of $\mathcal O(m_Z^2/v^2 y_t^2)$ lead to a Higgs boson mass reduction. This means that parameter points for which only the relaxed mass constraint is fulfilled, actually correspond to a Higgs boson mass below 122~GeV within the used approximation. Points with a relaxed mass constraint are especially those with small $\tan \beta$ and in particular all points with $\tan \beta \leq 5$. We know that other approximations including two-loop order terms lead to Higgs boson mass values within the interval $122~\text{GeV}\le m_h\le 129~\text{GeV}$ for $\tan \beta = 5$ and we observed that the deviations are of similar size independent on whether the $\mathcal O(m_Z^2/v^2 y_t^2)$ terms are taken into account or not (compare with Fig.~\ref{fig:hhhMSSM1l} right in the appendix). Therefore we include also the points with a relaxed mass constraint.

In Fig.~\ref{fig:hhhMSSM2lfull} the red points ({\color{red}  $\boldsymbol +$}) indicate that for this parameter point several Higgs bosons are expected to be discovered at the LHC while all the other points correspond to the single Higgs boson discovery region. The colours lightblue ({\color{lightblue}$\blacksquare$}), yellow ({\color{yellow}$\bullet$}) and green ({\color{green}$\boldsymbol \times$}) indicate the mass of the lighter top squark as below 1~TeV, between 1~TeV and 2.5~TeV and above 2.5~TeV, respectively. 

The largest deviations are found for $m_A = 200$~GeV and $\tan \beta = 5$ and amount to roughly $-15$~\%, with slightly larger deviations for the light stop mass in the interval $1.0~\text{TeV} \le m_{\tilde{t}_1} < 2.5$~TeV than for the light stop heavier than 2.5~TeV. 
Taking only points which fulfill the Higgs mass constraint $122~\text{GeV}\le m_h\le 129~\text{GeV}$ within the applied approximation leads to 
a maximal deviation of 
$-3$~to~$-4$~\% for $m_A \approx 240$~GeV, $\tan\beta \approx 7$ and the lighter stop mass in the interval $1.0~\text{TeV} \le m_{\tilde{t}_1} < 2.5$~TeV; for the light stop being heavier than 2.5~TeV the maximal deviations in this case are about $-2$~\% for $m_A \approx 280$~GeV
 and  $\tan\beta \approx 8$.

Including the $y_b$ terms does not have a substantial impact on the allowed deviations. In Fig.~\ref{fig:hhhMSSM2lfull} only points with a $\Delta \rho$ contribution from stops and bottoms (see~\cite{Deltarho}) smaller than 0.001 are included~\cite{ErlerLangackerPDG:2012}  -- however this constraint does not change the result qualitatively given that direct constraints are generally more constraining for superpartners than precision electroweak constraints. 

In summary, we have seen that the answer to the question of how large deviations of the triple Higgs coupling can be in the MSSM is quite sensitive to the applied approximation. For large $m_A$, the decoupling effect leads to small deviations, below $\sim -2$~\%, while for $m_A \approx 200$~GeV and small $\tan\beta$  the deviations can be as high as order $\mathcal O(-15\%)$. This sets the target for the self-coupling measurement in the MSSM context.

\subsection{Next to Minimal Supersymmetric Standard Model}

We now consider triple Higgs coupling deviations in the
NMSSM~\cite{Ellwanger:2009dp}. In this section we will assume that the
CP-even singlet in the NMSSM is very heavy and thus undetectable at the
LHC. If the singlet soft mass, $m_S$,  is made very large (while other soft
parameters are not changed), the NMSSM Higgs sector reduces to a MSSM-like
two Higgs doublet model and a decoupled singlet\footnote{This can be
checked for instance by looking at the scalar and pseudoscalar mass
matrices given in Ref.~\cite{Ellwanger:2009dp}, where  the most general
superpotential has been considered, and taking the limit $m_S \to
\infty$.}. If $m_A$ and $\tan \beta$ are taken as input parameters,
effectively this limit amounts to taking an additional term in the scalar
potential,
\beq
\Delta V =\lambda_S^2 (H_d^{0})^2 (H_u^{0})^2
\eeq
where $H_u^0$ and $H_d^0$ are the neutral components of the up- and down-type Higgs doublets. This leads to a contribution to the triple Higgs vertex\footnote{One can derive triple Higgs vertex including  singlet-doublet mixing effects, in a similar way, by using  the effective potential given in Ref.~\cite{Gupta:2012fy}, which includes corrections due to singlet-doublet mixing. For our investigation, this would require a more detailed study of the discovery potential of the singlet-like Higgs boson at the LHC in order to find the target in this model. This, however, goes beyond the scope of this paper.},
\bea
\Delta g_{hhh}&=& -\frac{3v}{2}\lambda_S^2 \sin {2 \alpha} \cos(\alpha+ \beta),
\eea
in addition to the MSSM tree-level contribution. Once again in the decoupling limit we have $m_A \to \infty$ and $\alpha \to \beta - \frac{\pi}{2}$ so that $g_{hhh}$ again reduces to its SM value $3 m_h^2/v$ where, for the NMSSM in the decoupling limit,  we have 
\beq
m_h^2= m_Z^2 \cos^2 2 \beta +\frac{1}{2}\lambda_S^2 v^2 \sin^2 2 \beta +\Delta_t,
\label{nmssmMass}
\eeq
where $\Delta_t$ is the loop contribution due to the stops discussed in detail in the last subsection.

\begin{figure}
\begin{center}
\includegraphics[width=0.8\columnwidth]{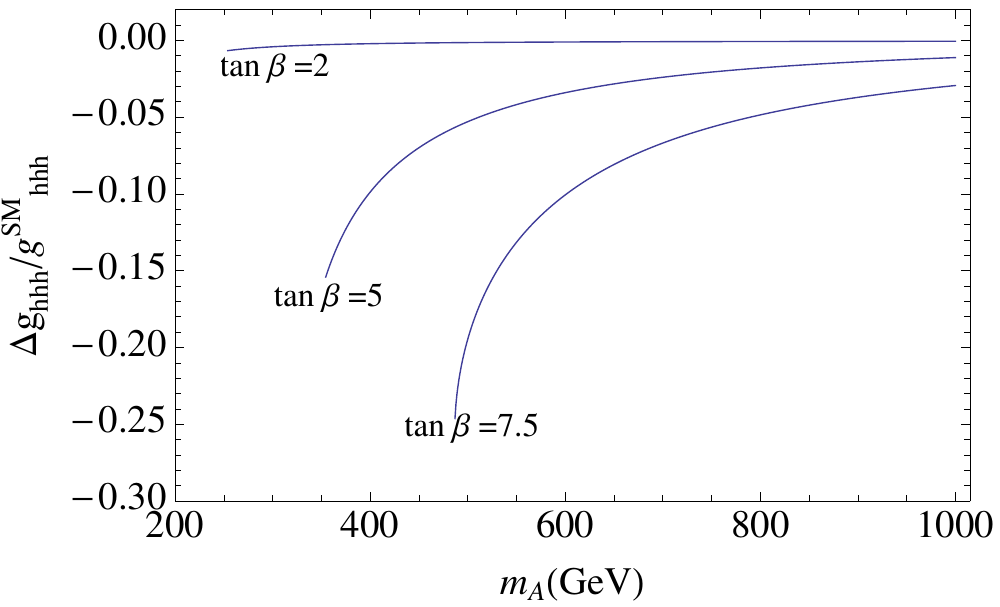}
\end{center}
\caption{The triple Higgs coupling target in the NMSSM with $M_S=500\gev$ as a function of $m_A$ for  various values of $\tan \beta$ where the singlet coupling remains perturbative below $10\tev$. }
\label{nmssm}
\end{figure}
In Fig.~\ref{nmssm} we show the Higgs self-coupling deviation as a function of $m_A$ for $\tan \beta =2,5$ and 7.5. 
We have assumed $M_S=500$ GeV and $X_t=0$ for the stop loop  contribution\footnote{For higher stop masses, lower values of $\lambda_S$ would have to be chosen for the same Higgs boson mass and hence the model would be more MSSM-like.}. The self-coupling deviation increases with $\tan \beta$ because the value of $\lambda_S$ required to raise the Higgs mass to 126 GeV increases (this can be understood from the $ \sin^2 2 \beta$ factor in eq.~Ê\ref{nmssmMass}).  For $\tan \beta =2$ we find $\lambda_S\leq0.7$, which satisfies the condition for perturbativity up to  the  Grand Unification scale~\cite{Hardy:2012ef} ($M_{GUT}\sim 2\times 10^{16}\gev$), whereas for  $\tan \beta =7.5$ we find  $\lambda_S\leq 2$, the upper value ($\lambda_S=2$) leading to a divergence in $\lambda_S$ at $\sim 10\tev$~\cite{Hall:2011aa}. 
For $\tan \beta > 7.5$ we find that the condition for perturbativity up to 10 TeV,  $\lambda_S <2$, is
not satisfied.
 Thus the maximum possible deviation, if we require   perturbativity up to  10 TeV is about   $-25 \%$ for $\tan \beta =7.5, m_A= 500$ GeV. 

Now we come to the question, would the heavier Higgs remain undetected by the LHC for this point $\tan \beta =7.5, m_A= 500$ GeV? In the case of the MSSM this point lies outside the LHC reach of heavy supersymmetric Higgs searches (see Fig.~1.21 of Ref.~\cite{Linssen:2012hp}). In the NMSSM  the coupling of the heavier Higgs bosons to down-type quarks and vector bosons is the same up to the percent level while the coupling  to up-type quarks  is reduced with respect to the MSSM. This means that the we expect similar (in processes controlled by heavy Higgs boson couplings to down-type fermions like $bb \to H \to \tau \tau$ ) or smaller cross-sections (if the process involves, for instance, gluon fusion where coupling to the top would be suppressed relative to the MSSM). Thus we would expect that if a point like $\tan \beta =7.5, m_A= 500$ GeV is beyond LHC reach for the MSSM the same would hold for the NMSSM too, given our construction. Thus  $\tan \beta =7.5, m_A= 500$ GeV indeed represents a point where the self-coupling deviation from SM is maximal, and the heavy Higgs bosons are beyond the LHC reach.  The self-coupling deviation for this point, $-25 \%$ is thus the target in the case of the NMSSM.

\section{Conclusions}

\begin{table}[t]
\begin{center}
\begin{tabular}{lc}
\hline\hline 
     Model  & $\Delta g_{hhh}/g^{SM}_{hhh}$  \\
\hline
Mixed-in Singlet & $-18$\,\%  \\
Composite Higgs & tens of \%  \\
Minimal Supersymmetry & $-2\,\%^a$~~$-15\,\%^b$  \\ 
NMSSM & $-25\,\%$ \\
LHC 3 ab$^{-1}$~\cite{Goertz:2013kp} & $[-20\,\%, +30\,\%]$ \\
\hline\hline
\end{tabular}
\caption{\label{table:results} Summary of the physics-based targets for the triple Higgs boson coupling. The target is based on scenarios where no other exotic electroweak symmetry breaking state (e.g., new Higgs bosons or ``$\rho$ particle") is found at the LHC except one: the $\sim 126\gev$ SM-like Higgs boson. 
Percentages quoted are approximate maximal deviations for each model based on the discussion in the text. 
For the $\Delta g_{hhh}/g^{SM}_{hhh}$ values of supersymmetry, superscript $a$ refers to the case of high $\tan\beta>10$ { and} no superpartners are found at the LHC, and superscript $b$ refers to all other cases, 
with the maximum  value of $-15\%$ reached for the special case of $\tan\beta\simeq 5$. In the last row, the best estimates for the $1\sigma$ accuracy of the measurement of the triple Higgs coupling at the LHC with 3 ab$^{-1}$ integrated luminosity is given. It is assumed here that no additional dynamics or operators contribute to non-SM shifts in $pp\to hh$ except the self-coupling.}
\end{center}
\vspace{-0.6cm}
\label{sum}
\end{table}

To summarize, we have found that the $150\mev$ uncertainty on the Higgs boson mass that ATLAS and CMS are scheduled to achieve is likely to be better than we will ever need to know it 
in the foreseeable future. Better determinations yield no obvious advantage in testing any proposed question about nature that we can formulate today. 

On the other hand, we have shown that in beyond the SM scenarios there can be significant gains in our understanding of nature -- including discovering and discerning new physics -- if the self-coupling measurement can be performed much more accurately than LHC projections afford. Generally speaking, measurements need to determine the Higgs self-coupling to better than 20\% in order to have the chance to see effects of new physics in this important channel, if no other dynamics associated with electroweak symmetry breaking are seen. Our results for different BSM theories are summarized in Table~Ê\ref{sum}. This is a challenging target for future LHC upgrades and proposed $e^+e^-$ linear collider options of the future.


\appendix
\addcontentsline{toc}{section}{Appendix}
\section*{Appendix} 

The result for the tri-Higgs coupling presented in sec.~\ref{sec:mssm} turns out to be sensitive to the choice of higher-order terms included in the computation. It is conceivable that adding more terms self-consistently to the tri-Higgs coupling and mass, and re-computing would also yield a change of similar size. For this reason we have including this appendix to show the changes in the size of the tri-Higgs coupling for various approximations that exist in the literature.

We start our discussion taking into account radiative corrections to the Higgs triple coupling of the order of $m_t^4$. In the literature, two approximations can be found: one including all $m_t^4$ contributions \cite{mt4,Yukmt4hhh,Williams:2011bu} and the other one based on a renormalization-group improved leading-log approximation~\cite{llog}. In the latter approach also two-loop leading-log terms are provided which are taken into account in sec.~\ref{sec:mssm}.

In the leading-log approach we evaluate the coefficients $\lambda_i$ with $i = 1,...,7$ of the effective potential eq.~\eqref{VeffMSSM}, given in 
Ref.~\cite{llog} as (showing here explicitly only the tree-level part and the $m_t^4$ one-loop contributions)
{\allowdisplaybreaks
\begin{align}\label{la1}
\lambda_1 &= \frac{m_Z^2}{v^2} - \frac{y_t^4}{32 \pi^2}\frac{\mu^4}{M_{\text{SUSY}}^4}\\
\lambda_2 &= \frac{m_Z^2}{v^2}  
+ \frac{3 y_t^4}{8 \pi^2} 
\left[\log{\frac{m_t^2}{M_{\text{SUSY}}^2}} - \frac{A_t^2}{M_{\text{SUSY}}^2}\left(1 - \frac{A_t^2}{12 M_{\text{SUSY}}^2}\right)\right]\\
\lambda_3 &= -\frac{m_Z^2}{v^2}\left(1 - 2 c_W^2\right)
 + \frac{y_t^4}{32 \pi^2} \frac{\mu^2}{M_{\text{SUSY}}^2}\left(3 -  \frac{A_t^2}{M_{\text{SUSY}}^2}\right)\\
\lambda_4 &=  - 2 \frac{m_Z^2}{v^2}c_W^2 + \frac{y_t^4}{32 \pi^2} \frac{\mu^2}{M_{\text{SUSY}}^2}\left(3 -  \frac{A_t^2}{M_{\text{SUSY}}^2}\right)\\
\lambda_5 &= - \frac{y_t^4}{32 \pi^2}\frac{\mu^2 A_t^2}{M_{\text{SUSY}}^4}\\
\lambda_6 &= - \frac{y_t^4}{32 \pi^2}\frac{\mu^3 A_t}{M_{\text{SUSY}}^4}\\
\lambda_7 &=   \frac{y_t^4}{32 \pi^2}\frac{\mu}{M_{\text{SUSY}}}\left(\frac{A_t^3}{M_{\text{SUSY}}^3}- 
6\frac{A_t}{M_{\text{SUSY}}}\right)\label{la7}
\end{align}
with $c_W = m_W/m_Z$ being the cosine of the weak mixing angle and $m_W$ the $W$~mass. The parameter $M_{\text{SUSY}}^2$ is defined as the average of the stop masses squared, $M_{\text{SUSY}}^2 = (m_{\tilde{t}_1}^2 + m_{\tilde{t}_2}^2)/2$. The top soft breaking trilinear coupling is denoted as $A_t$ and the top Yukawa coupling as $y_t$.

\begin{figure*}
\begin{center}
\includegraphics[width=0.9\columnwidth]{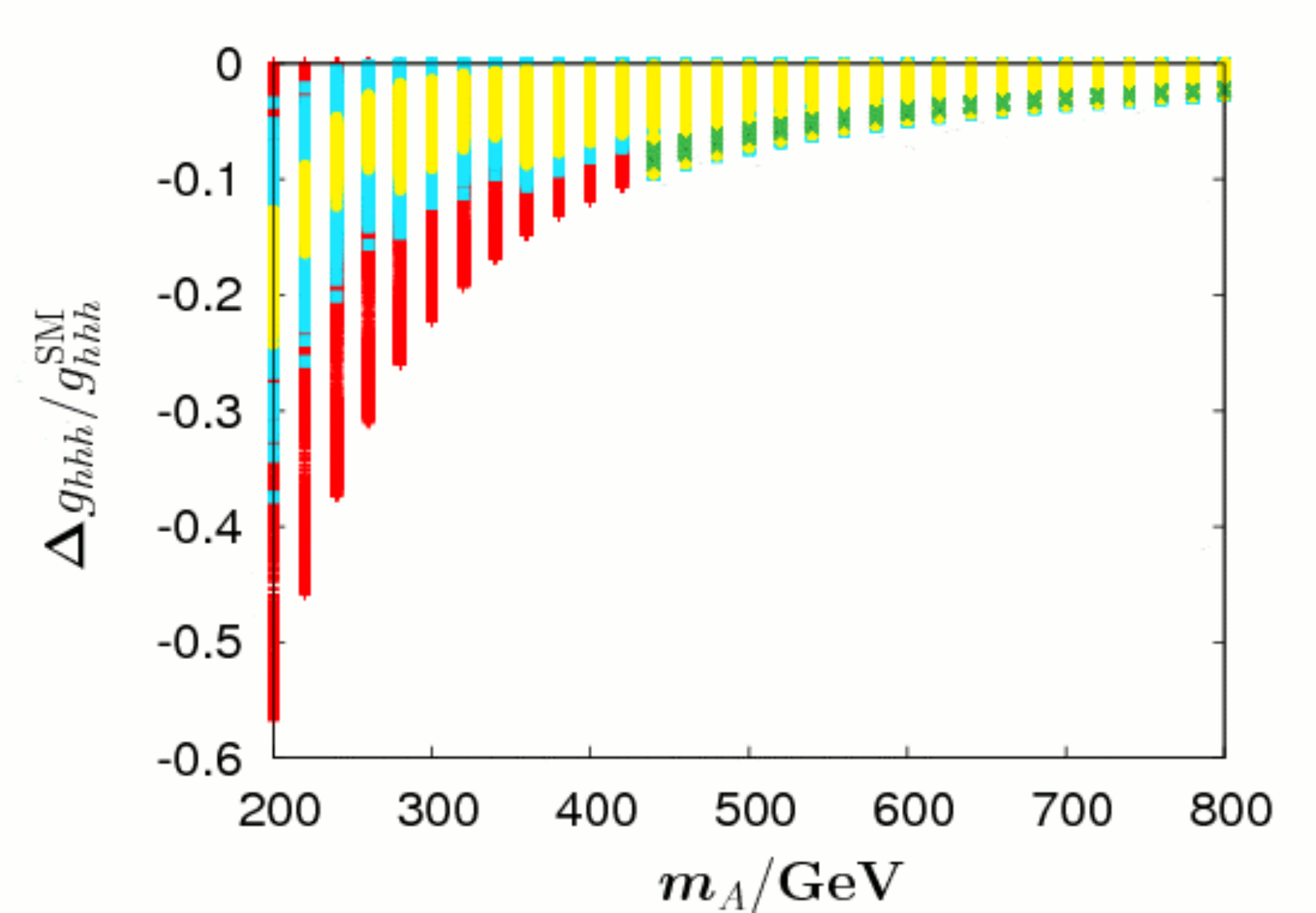}
\includegraphics[width=0.9\columnwidth]{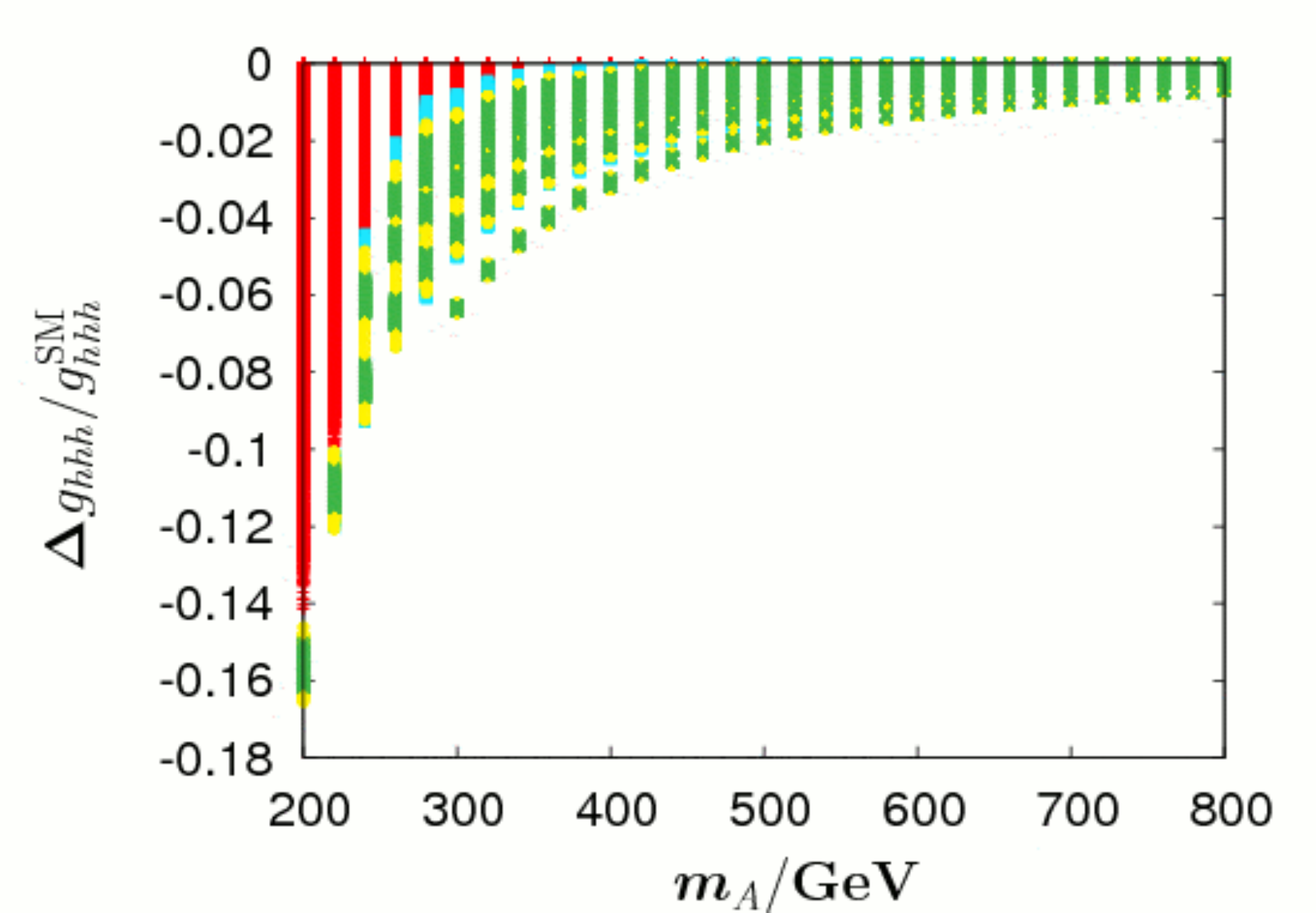}
\end{center}
\caption{The triple Higgs coupling deviations in the MSSM as a function of $m_A$ at $\mathcal O(y_t^4)$ in the renormalization-group improved leading-log approximation, with $y_t$ defined via the on-shell top quark mass, $y_t = \sqrt{2} m_t/(v \sin \beta)$ with $m_t = 173.2$~GeV (left plot) and
including additionally $O(y_t^4 \alpha_s)$ and $O(y_t^6)$ terms, with $y_t$ defined via the running mass $y_t = \sqrt{2} \overline{m}_t/(v \sin \beta)$ (right plot). 
 The red points ({\color{red}  $\boldsymbol +$}) correspond to parameter points for which several Higgs bosons can be discovered at the LHC. All other points belong to the single Higgs boson discovery region and are coloured according to the mass value of the lighter top squark $m_{\tilde{t}_1}$: 
lightblue ({\color{lightblue}$\blacksquare$}) for $m_{\tilde{t}_1} < 1.0$~TeV, yellow ({\color{yellow}$\bullet$}) for $1.0 \leq m_{\tilde{t}_1} < 2.5$~TeV, and green ({\color{green}$\boldsymbol \times$}) for $m_{\tilde{t}_1} \ge 2.5$~TeV.
}
\label{fig:hhhMSSM1l}
\end{figure*}

 In Fig.~\ref{fig:hhhMSSM1l} the relative deviations of the triple Higgs coupling,
\beq
\Delta g_{hhh}/g^{SM}_{hhh},\quad \Delta g_{hhh}= g_{hhh} - g^{SM}_{hhh}
\eeq
are shown versus $m_A$ with the MSSM coupling, $g_{hhh}$, as given in eq.~\eqref{ghhh}. The parameter scan is performed as described in 
sec.~\ref{sec:mssm}.  Only points with a Higgs boson mass of $122~\text{GeV}\leq m_h \leq 129~\text{GeV}$, evaluated in the particular 
approximations, are taken into account. 

In the left plot of Fig.~\ref{fig:hhhMSSM1l} the deviations are calculated taking only into account the one-loop $m_t^4$ terms, see  eqs.~\eqref{la1}--\eqref{la7}. The top Yukawa coupling is chosen as $y_t = \sqrt{2} m_t/(v \sin \beta)$ with the top quark mass $m_t = 173.2$~GeV. The red region ({\color{red} $\boldsymbol +$}) corresponds to parameter points in the $m_A - \tan \beta$ plane with a several Higgs boson discovery potential at the LHC according to Fig.~1.21 of Ref.~\cite{Linssen:2012hp}. All other points belong to the region where only a single Higgs boson can be discovered. The largest deviations can be found for low $m_A$ and can be larger than 30\% in this approximation. The colour coding of the single Higgs boson discovery region is the following: 
lightblue ({\color{lightblue}$\blacksquare$}) for the mass of the lighter top squark $m_{\tilde{t}_1}$ below 1~TeV, 
 yellow ({\color{yellow}$\bullet$}) for $m_{\tilde{t}_1}$ between 1.0~and~2.5~TeV,  and green ({\color{green}$\boldsymbol \times$}) for $m_{\tilde{t}_1}$ larger than 2.5~TeV. 
Within this leading log approximation the largest deviations occur only for rather light stop masses, only for large $m_A$ we find points corresponding to stop masses larger than 2.5~TeV. This is mainly due to the fact that the one-loop corrections to the Higgs mass are very large and for large stop masses the light Higgs boson becomes too heavy with mass values above 129~GeV. The two-loop corrections to the Higgs boson mass decrease the mass so that it is to expect that at two-loop order for a wider range of deviations large stop masses can be found. 

Using the exact $m_t^4$ approximation for $g_{hhh}$ as given in Ref.~\cite{mt4,Yukmt4hhh,Williams:2011bu} gives a similar picture if the same approximation is used in the SM. The complete $m_t^4$ contributions include a constant term which is also present in the SM,
\begin{align}
g_{hhh}^{\text{SM}} = \frac{3}{v} \left(m^2_h - \frac{\sqrt{2} G_F }{\pi^2}{m_t^4}\right)
\end{align}
and which has to be taken into account to ensure a proper decoupling \cite{Yukmt4hhh,Dobado:2002jz} ($m_h$ is also understood to be calculated in the complete $m_t^4$ approximation, of course). We find slight positive deviations, below 1\%, in the complete $m_t^4$ approximation which are not present in the renormalization group improved result, which is a rather small effect. However, going to larger $X_t$ values, larger than the ones chosen in the scan we find rather big deviations and the approximations seem to not work properly anymore.

In the left plot of Fig.~\ref{fig:hhhMSSM1l} the deviations have been calculated with the top Yukawa coupling expressed by the top quark pole
 mass. Applying instead  
$y_t = \sqrt{2} \overline{m}_t/(v \sin \beta)$ with the running top quark mass 
$\overline{m}_t =  m_t/[1+4 \alpha_s/(3 \pi)]$, as used in Ref.~\cite{llog},  leads to a 
rescaling of the one-loop contributions by $\overline{m}_t^4/m_t^4$. 
As $\overline{m}_t < m_t$ this decreases the one-loop contributions to the triple Higgs coupling as well as to the Higgs 
mass. As a  result the size  of the couplings deviations are reduced, for $m_A = 200$~GeV from roughly 0.35 to 0.25, while
the regions, where large stop masses can be found, are enlarged. 

In Fig.~\ref{fig:hhhMSSM1l} right, two-loop contributions to the effective potential of the order $\mathcal O(y_t^4 \alpha_s)$ and $\mathcal O(y_t^6)$ have been taken into account. The top Yukawa coupling again is evaluated with the running top quark mass. The colour coding is the same as before. In Fig.~\ref{fig:hhhMSSM1l} the deviations of the couplings are shown in dependence on $m_A$. A further decrease of the deviations and an increase of the region with very large stop masses can be observed. Stop mass values larger than 2.5~TeV cover almost the complete single Higgs boson discovery region. 

These results can be compared with the results shown in sec.~\ref{sec:mssm} Fig.~\ref{fig:hhhMSSM2lfull}. Taking into account the complete 
set of corrections given in 
Ref.~\cite{llog} particularly reduces the Higgs mass and therefore, keeping strictly the mass constraint of 
$122~\text{GeV}\leq m_h \leq 129~\text{GeV}$ 
leads to a sizeable reduction of the possible deviations as no parameter points are found for $\tan \beta = 5$ and hence no points for $m_A = 200$~GeV in the single Higgs boson discovery region, for which before we found the largest 
deviations. Relaxing however the mass constraint leads to a similar size of the deviations (see sec.~\ref{sec:mssm} and Fig.~\ref{fig:hhhMSSM2lfull}).



\begin{thebibliography}{90}
\baselineskip=12.05pt

\bibitem{HiggsDiscovery}
ATLAS Collaboration,
  Phys.\ Lett.\ B {716} (2012) 1 [arXiv:1207.7214].
CMS Collaboration,
  Phys.\ Lett.\ B {716} (2012) 30 [arXiv:1207.7235].

\bibitem{DiHiggs}
See e.g.\  J.~Baglio, A.~Djouadi, R.~Gr\"ober, M.~M.~M\"uhlleitner, J.~Quevillon and M.~Spira,
  JHEP {\bf 1304} (2013) 151
  [arXiv:1212.5581 [hep-ph]];
M.~J.~Dolan, C.~Englert and M.~Spannowsky,
  JHEP {\bf 1210} (2012) 112
  arXiv:1206.5001 [hep-ph];
  Phys.\ Rev.\ D {\bf 87} (2013) 055002
  arXiv:1210.8166 [hep-ph];
M.~Gillioz, R.~Gr\"ober, C.~Grojean, M.~M\"uhlleitner and E.~Salvioni,
  JHEP {\bf 1210}, 004 (2012)
  [arXiv:1206.7120 [hep-ph]];
M.~Gouzevitch, A.~Oliveira, J.~Rojo, R.~Rosenfeld, G.~Salam and V.~Sanz,
  arXiv:1303.6636 [hep-ph];
J.~Cao, Z.~Heng, L.~Shang, P.~Wan and J.~M.~Yang,
  JHEP {\bf 1304}, 134 (2013)
  [arXiv:1301.6437 [hep-ph]].
   
\bibitem{Gupta:2012mi} 
  R.~S.~Gupta, H.~Rzehak and J.~D.~Wells,
  arXiv:1206.3560 [hep-ph].
  
\bibitem{Ball:2007zza} 
  G.~L.~Bayatian {\it et al.}  [CMS Collaboration],
  ``CMS technical design report, volume II: Physics performance,''
  J.\ Phys.\ G {\bf 34}, 995 (2007).
  
\bibitem{ATLASTDR}
ATLAS Collaboration,  ``ATLAS: Detector and physics performance technical design report. Volume 2,''
  CERN-LHCC-99-15 (May 1999).
  
\bibitem{Holthausen:2011aa} 
For other similar considerations, see for example,  M.~Holthausen, K.~S.~Lim and M.~Lindner,
  JHEP {\bf 1202}, 037 (2012)
  [arXiv:1112.2415 [hep-ph]].
  
\bibitem{Wells:2009kq} 
  J.~D.~Wells,
  arXiv:0909.4541 [hep-ph].
  
\bibitem{Degrassi:2012ry} 
  G.~Degrassi, S.~Di Vita, J.~Elias-Miro, J.~R.~Espinosa, G.~F.~Giudice, G.~Isidori, A.~Strumia and ,
  JHEP {\bf 1208}, 098 (2012)
  [arXiv:1205.6497 [hep-ph]].
  
\bibitem{Masina:2012tz} 
  I.~Masina,
  arXiv:1209.0393 [hep-ph].
  
\bibitem{Tobe:2002zj} 
  K.~Tobe and J.~D.~Wells,
  Phys.\ Rev.\ D {\bf 66}, 013010 (2002)
  [hep-ph/0204196].

  \bibitem{Giudice:2011cg} 
  G.~F.~Giudice and A.~Strumia,
  Nucl.\ Phys.\ B {\bf 858}, 63 (2012)
  [arXiv:1108.6077 [hep-ph]].


\bibitem{Peskin:2012we} 
  M.~E.~Peskin,
  arXiv:1207.2516 [hep-ph].

\bibitem{Beneke:2000hk} 
  M.~Beneke {\it et al.},
  In *Geneva 1999, Standard model physics (and more) at the LHC* 419-529
  [hep-ph/0003033].
    
\bibitem{AguilarSaavedra:2001rg} 
  J.~A.~Aguilar-Saavedra {\it et al.}  [ECFA/DESY LC Physics Working Group Collaboration],
  hep-ph/0106315.
 
\bibitem{Baur:2002qd} 
  U.~Baur, T.~Plehn and D.~L.~Rainwater,
  Phys.\ Rev.\ D {\bf 67}, 033003 (2003)
  [hep-ph/0211224];
  Phys.\ Rev.\ D {\bf 68}, 033001 (2003)
  [hep-ph/0304015].
 
\bibitem{Wang:2007zx} 
  L.~Wang, W.~Wang, J.~M.~Yang and H.~Zhang,
  Phys.\ Rev.\ D {\bf 76}, 017702 (2007)
  [arXiv:0705.3392 [hep-ph]].
  
  \bibitem{Wells}
  R.~Schabinger, J.~D.~Wells,
  Phys.\ Rev.\  {\bf D72}, 093007 (2005)
  [hep-ph/0509209];

\bibitem{Bowen:2007ia}
  M.~Bowen, Y.~Cui, J.~D.~Wells,
  JHEP {\bf 0703}, 036 (2007)
  [hep-ph/0701035].

\bibitem{hep-ph/9504378} 
  M.~Spira, A.~Djouadi, D.~Graudenz and P.~M.~Zerwas,
  Nucl.\ Phys.\ B\ {\bf 453}, 17  (1995)
  [hep-ph/9504378].

  \bibitem{hep-ph/9409380} 
  K.~Hagiwara, S.~Matsumoto, D.~Haidt and C.~S.~Kim,
  Z.\ Phys.\ C\ {\bf 64}, 559  (1994)
  [Erratum-ibid.\ C\ {\bf 68}, 352  (1995)]
  [hep-ph/9409380].

\bibitem{723875} 
  W.~M.~Yao {\it et al.} [Particle Data Group Collaboration],
  J.\ Phys.\ GG\ {\bf 33}, 1  (2006).

\bibitem{Djouadi:2005gj} 
  A.~Djouadi,
  Phys.\ Rept.\  {\bf 459}, 1 (2008)
  [hep-ph/0503173].
  
\bibitem{Hardy:2012ef} 
For a description of how larger $\lambda_S$ can be compatible with unification see,  E.~Hardy, J.~March-Russell and J.~Unwin,
ÊÊJHEP {\bf 1210}, 072 (2012)
ÊÊ[arXiv:1207.1435 [hep-ph]].
ÊÊ
  
\bibitem{Linssen:2012hp} 
  L.~Linssen, A.~Miyamoto, M.~Stanitzki and H.~Weerts,
  arXiv:1202.5940 [physics.ins-det].

 \bibitem{ArkaniHamed:2002qy} 
  N.~Arkani-Hamed, A.~G.~Cohen, E.~Katz and A.~E.~Nelson,
  JHEP {\bf 0207}, 034 (2002)
  [hep-ph/0206021].
  
\bibitem{Agashe:2004rs} 
  K.~Agashe, R.~Contino and A.~Pomarol,
  Nucl.\ Phys.\ B {\bf 719}, 165 (2005)
  [hep-ph/0412089].
  
  \bibitem{Giudice:2007fh} 
  G.~F.~Giudice, C.~Grojean, A.~Pomarol and R.~Rattazzi,
  JHEP {\bf 0706}, 045 (2007)
  [hep-ph/0703164].
  
\bibitem{Pappadopulo:2013vca} 
  D.~Pappadopulo, A.~Thamm and R.~Torre,
  arXiv:1303.3062 [hep-ph].
    
  \bibitem{Manohar:1983md} 
  A.~Manohar and H.~Georgi,
  Nucl.\ Phys.\ B {\bf 234}, 189 (1984).

\bibitem{Delaunay:2007wb} 
  C.~Delaunay, C.~Grojean and J.~D.~Wells,
  JHEP {\bf 0804}, 029 (2008)
  [arXiv:0711.2511 [hep-ph]].


 
 
\bibitem{Carena:2000uj} 
 M.~S.~Carena, D.~Garcia, U.~Nierste and C.~E.~M.~Wagner,
  Nucl.\ Phys.\ B {\bf 577} (2000) 88
  [hep-ph/9912516].

\bibitem{ErlerLangackerPDG:2012}
J. Erler, P. Langacker, ``Electroweak Model and Constraints on New Physics,"
from J. Beringer et al., Phys.\ Rev.\ D86, 010001 (2012). http://pdg.lbl.gov (June 18, 2012).

\bibitem{llog}M.~S.~Carena, J.~R.~Espinosa, M.~Quiros and C.~E.~M.~Wagner,
  Phys.\ Lett.\ B {\bf 355} (1995) 209
  [hep-ph/9504316].

\bibitem{HaHe} H.~E.~Haber and R.~Hempfling,
  Phys.\ Rev.\ D {\bf 48} (1993) 4280
  [hep-ph/9307201].


\bibitem{Deltarho}
M.~Drees and K.~Hagiwara,
  Phys.\ Rev.\ D {\bf 42} (1990) 1709.
 
  
  
\bibitem{Goertz:2013kp} 
  F.~Goertz, A.~Papaefstathiou, L.~L.~Yang and J.~Zurita,
  arXiv:1301.3492 [hep-ph].
  

\bibitem{Ellwanger:2009dp} 
 U.~Ellwanger, C.~Hugonie and A.~M.~Teixeira,
   Phys.\ Rept. {\bf 496}, 1 (2010)
  [arXiv:0910.1785 [hep-ph]].

\bibitem{Gupta:2012fy} 
  R.~S.~Gupta, M.~Montull and F.~Riva,
  arXiv:1212.5240 [hep-ph].

\bibitem{Hall:2011aa} 
  L.~J.~Hall, D.~Pinner and J.~T.~Ruderman,
  JHEP {\bf 1204}, 131 (2012)
  [arXiv:1112.2703 [hep-ph]].


\bibitem{mt4} V.~D.~Barger, M.~S.~Berger, A.~L.~Stange and R.~J.~N.~Phillips,
  Phys.\ Rev.\ D {\bf 45} (1992) 4128.


\bibitem{Yukmt4hhh}  W.~Hollik and S.~Penaranda,
  Eur.\ Phys.\ J.\ C {\bf 23} (2002) 163
  [hep-ph/0108245].

\bibitem{Williams:2011bu}
  K.~E.~Williams, H.~Rzehak and G.~Weiglein,
  Eur.\ Phys.\ J.\ C {\bf 71} (2011) 1669
  [arXiv:1103.1335 [hep-ph]].

\bibitem{Dobado:2002jz}
  A.~Dobado, M.~J.~Herrero, W.~Hollik and S.~Penaranda,
  Phys.\ Rev.\ D {\bf 66} (2002) 095016
  [hep-ph/0208014].


\end{thebibliography}
\end{document}